\documentclass[12pt]{iopart}

\usepackage{graphicx}
\usepackage{subfig}
\usepackage{amssymb}
\usepackage{amsthm} 
\usepackage{mathptmx}
\usepackage{nicefrac}
\usepackage{xcolor}
\usepackage[switch]{lineno} 

\begin{document}

\title[]{First measurement of kaonic helium-4 M-series transitions}

\author{ F Sgaramella$^{1*}$, D Sirghi$^{2,1,3**}$, L Abbene$^4$, F Artibani$^1$, M Bazzi$^1$, D Bosnar$^6$, M Bragadireanu$^3$, A Buttacavoli$^4$, M Cargnelli$^5$, M Carminati$^{7,8}$, A Clozza$^1$, F Clozza$^1$, G Deda$^{7,8}$, R Del Grande$^{9,1}$, L De Paolis$^1$, K Dulski$^{1,11,12}$, L Fabbietti$^9$, C Fiorini$^{7,8}$, I Fri\v{s}\v{c}i\'c$^6$, C Guaraldo$^1$, M Iliescu$^1$, M Iwasaki$^{10}$, A Khreptak$^{1,11}$, S Manti$^1$, J Marton$^5$, M Miliucci$^{1,\dagger}$, P Moskal$^{11,12}$, F Napolitano$^1$, S Nied\'{z}wiecki$^{11,12}$, H Ohnishi$^{13}$, K Piscicchia$^{3,1}$, F Principato$^4$, A Scordo$^{1}$, M Silarski$^{11}$, F Sirghi$^{1,3}$, M Skurzok$^{11,12}$, A Spallone$^1$, K Toho$^{13}$, M T\"uchler$^5$, O Vazquez Doce$^1$, C Yoshida$^{13}$, J Zmeskal$^5$ and C Curceanu$^1$}

\address{$^1$ Laboratori Nazionali di Frascati INFN Frascati, Italy}
\address{$^2$ Centro Ricerche Enrico Fermi – Museo Storico della Fisica e Centro Studi e Ricerche “Enrico Fermi”, Roma, Italy}
\address{$^3$ Horia Hulubei National Institute of Physics and Nuclear Engineering (IFIN-HH) Măgurele, Romania}
\address{$^4$ Department of Physics and Chemistry (DiFC)—Emilio Segrè, University of Palermo, Palermo, Italy}
\address{$^5$ Stefan-Meyer-Institut f\"ur Subatomare Physik, Vienna, Austria}
\address{$^6$ Department of Physics, Faculty of Science, University of Zagreb, Zagreb, Croatia}
\address{$^7$ Politecnico di Milano, Dipartimento di Elettronica, Informazione e Bioingegneria, Milano, Italy}
\address{$^8$ INFN Sezione di Milano, Milano, Italy}
\address{$^9$ Excellence Cluster Universe, Technische Universiät München Garching, Germany}
\address{$^{10}$ RIKEN, Tokyo, Japan}
\address{$^{11}$ Faculty of Physics, Astronomy, and Applied Computer Science, Jagiellonian University, Krak\'{o}w, Poland}
\address{$^{12}$ Center for Theranostics, Jagiellonian University, Krakow, Poland}
\address{$^{13}$ Research Center for Electron Photon Science (ELPH), Tohoku University, Sendai, Japan}
\address{$^\dagger$ current address: ASI, Agenzia Spaziale Italiana, Roma, Italy }

\ead{$^*$ francesco.sgaramella@lnf.infn.it (Corresponding Author)}
\ead{$^{**}$ diana.sirghi@lnf.infn.it (Corresponding Author)}

\vspace{10pt}
\begin{indented}
\item[] September 2023
\end{indented}

\begin{abstract}
In this paper we present the results of a new kaonic helium-4 measurement with a 1.37 g/l gaseous target by the SIDDHARTA-2 experiment at the DA$\Phi$NE collider. We measured, for the first time, the energies and yields of three transitions belonging to the M-series. Moreover, we improved by a factor about three, the statistical precision of the 2p level energy shift and width induced by the strong interaction, obtaining the most precise measurement for gaseous kaonic helium, and measured the yield of the L$_\alpha$ transition at the employed density, providing a new experimental input to investigate the density dependence of kaonic atoms transitions yield.
\end{abstract}

%
\vspace{2pc}
\noindent{\it Keywords}: Kaonic Helium, X-rays spectroscopy, atomic cascade, kaon-nucleon interaction
%
\submitto{\jpg}
%
%
%
%

\section*{Introduction}
The study of kaonic atoms through X-ray spectroscopy provides valuable experimental data to probe the strong interaction between negatively charged kaons (K$^-$) and nuclei at low energy. This technique allows to extract information about the strong interaction between the kaon and the nucleus at threshold energy, by analyzing the shift and the broadening of the energy levels with respect to the values predicted by quantum electrodynamics (QED). Such data are crucial to constrain the quantum chromodynamics models in the non-perturbative regime with strangeness \cite{Lee:1994my,Bernard:2007zu,Cieply:2016jby,revmodphys:2019}. \\
A kaonic atom is formed when a K$^-$, with sufficiently low momentum, is stopped in a target and captured by an atom via the electromagnetic interaction. The capture occurs in highly excited state, determined by the reduced mass of the system. After the capture, the kaonic atom experiences a series of de-excitation processes, including Coulomb de-excitation and external Auger emission \cite{Kalantari:2012zz}, which bring the kaon to the ground state. These processes are accompanied by the emission of radiation which, for the transitions to the lower-lying level, is in the X-ray domain. Not all the kaons reach the ground state, since other processes may heavily influence the cascade. Among these processes, for exotic hydrogen and, for some extent, for helium, the Stark mixing is quite important \cite{Ericson:1970ij}. Stark effect is responsible for a drastic reduction of the X-ray yields to lower levels when the target density increases. For this reason, in addition to the study of the strong interaction, kaonic atoms X-ray spectroscopy is a unique tool to investigate the de-excitation mechanisms that occur in kaonic atoms, by measuring the X-ray yields of various transitions. Since the competing processes become more prevalent with density, they play a significant role in determining the X-ray yields, and the experimental density dependence of the yields can serve as a test bench for various cascade models \cite{Kalantari:2012isg,Berezin:1970qw}.\\
In this context, a special role is played by kaonic helium. In the 1970s and 1980s, three different experiments \cite{Wiegand:1971zz,Batty:1979a,Baird:1983ub} observed a large energy shift on the kaonic helium-4 2p level induced by the strong interaction, results which were in contradiction with the theoretical expectations \cite{Batty:1997,Hirenzaki:2000}. The so-called kaonic “helium puzzle" was solved by the E570 experiment at KEK \cite{Okada:2007ky}, confirmed also by the SIDDHARTA experiment \cite{bazzi:2009,SIDDHARTA:2012rsv}. More recent and precise measurements have been performed at J-PARC by E62 \cite{J-PARCE62:2022qnt}, using a liquid helium target, and by the SIDDHARTA-2 collaboration, employing a gaseous one \cite{SIDDHARTA2:2022,Sirghi:2023scw}.\\
The study of kaonic helium is a topic of great interest \cite{Tatsuno:2016zgx}, concerning both the X-ray energies and yields. Koike and Akaishi \cite{Koike:1998uz} developed a cascade model for kaonic helium, which reproduced the experimental data measured at liquid helium-4 density, but failed to match the data from the SIDDHARTA experiment \cite{Bazzi:2014}, which included measurements at various gas densities. As a result, the dependence of the yields in kaonic helium across the whole density scale, from liquid to gas, is still an open issue, requiring experimental input. \\
In this work, we report the recent measurement of gaseous kaonic helium-4 performed by the SIDDHARTA-2 collaboration at the DA$\Phi$NE collider of INFN-LNF. For the first time, we successfully identified and measured three M-series transitions in kaonic helium-4 and we determined their X-ray yields. Additionally, we measured the X-ray yield for some L-series transitions at the density of 1.37 $\pm$ 0.07 g/l, providing new valuable information on kaonic helium cascade process. Finally, we performed a new measurement of the 2p level energy shift and width, with threefold improved statistical precision with respect to the previous measurements performed with gaseous helium, setting a new record for gaseous targets.

\section{The SIDDHARTA-2 experiment}
\begin{figure}[htbp]
\centering
\mbox{\includegraphics[width=10 cm]{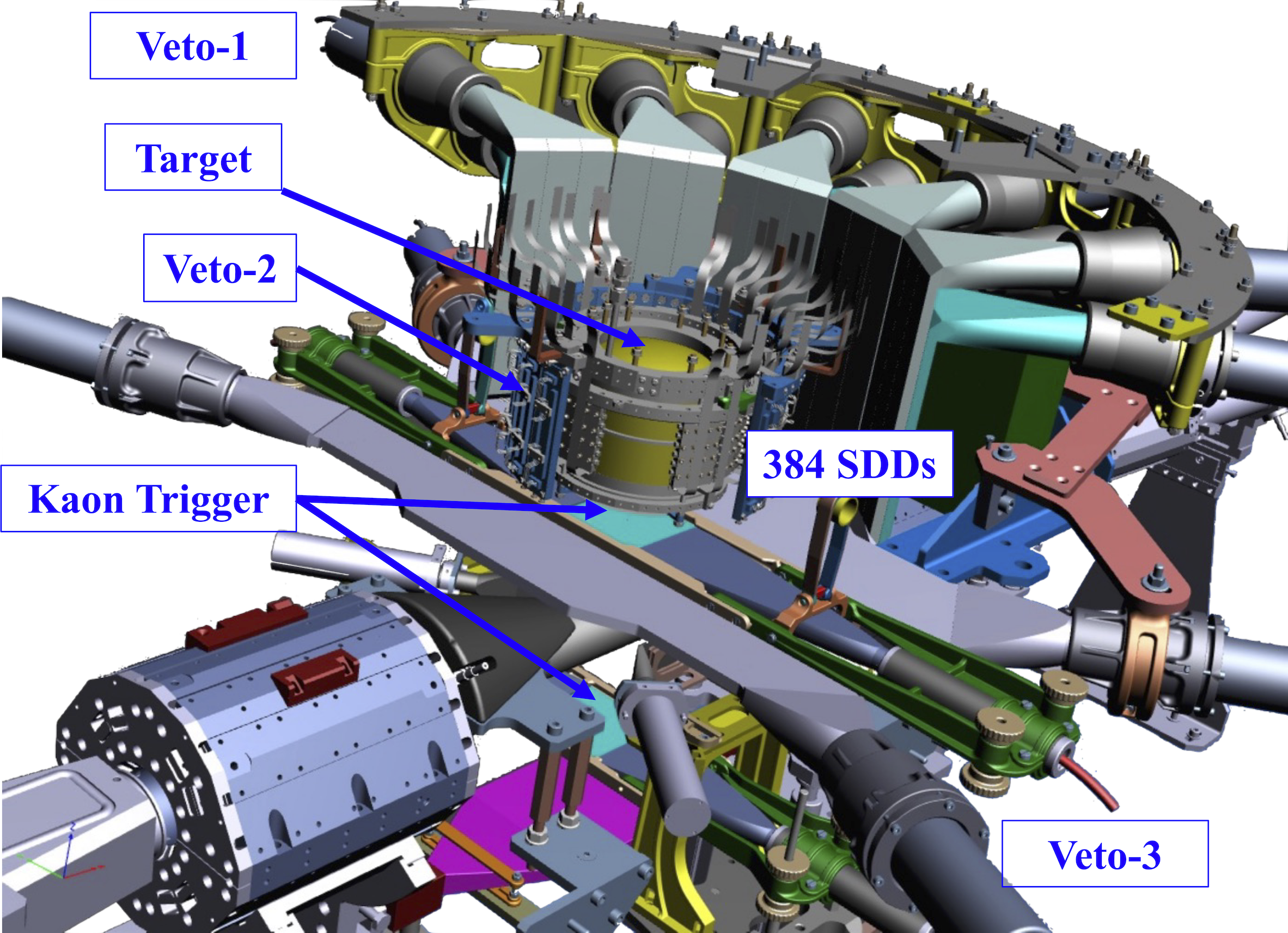}}
\caption{Schematic drawing of the SIDDHARTA-2 experimental apparatus with various elements of the setup indicated.}
\label{fig_setup}
\end{figure}

The SIDDHARTA-2 apparatus (see Figure \ref{fig_setup}) is installed above the interaction region (IR) of the DA$\Phi$NE collider \cite{Milardi:2018sih} at the National Institute of Nuclear Physics in Frascati (INFN-LNF). Low momentum (p = 127 MeV/c) and monochromatic ($\Delta$p/p = 0.1\%) kaons are delivered via the $\phi$-decay into a K$^+$K$^-$ pair.\\ The main goal of the experiment is to perform the first measurement of the strong interaction induced shift and width of the fundamental level in kaonic deuterium. This measurement, combined with the kaonic hydrogen one already performed by SIDDHARTA \cite{Bazzi:2011}, will allow extracting, for the first time, the experimental isospin dependent antikaon-nucleon scattering lengths \cite{revmodphys:2019}. To face the challenging kaonic deuterium measurement, the SIDDHARTA-2 collaboration developed a completely new apparatus with respect to the one used for the kaonic hydrogen measurement. The core of the setup consists of a cryogenic cylindrical target made of 150 $\mu$m kapton walls and a high purity aluminium frame to ensure an efficient cooling. The target can be filled with different types of gases. The cooling system permits to cool the gas down to 20 K, while the pressure can be tuned up to 1.4 bar to optimize the kaons' stopping efficiency and perform studies at different densities. The target is surrounded by 384 Silicon Drift Detectors (SDDs), covering an active area of 245.8 cm$^2$. The SDDs have been developed by Fondazione Bruno Kessler (FBK) in collaboration with INFN-LNF, Politecnico of Milano and the Stefan Meyer Institute (SMI), specifically for performing kaonic atoms measurements. The excellent energy and time resolutions (FWHM), 157.8$\pm$0.3 eV at 6.4 keV \cite{Miliucci:2021wbj} and 500 ns \cite{Miliucci:2022lvn}, respectively, are fundamental  for the background reduction and, consequently, the success of the measurement. Another factor of merit of the setup is the capability to determine the X-ray energy with a systematic error of a few eV \cite{Sgaramella:2022}, making SIDDHARTA-2 the ideal experiment to perform high precision kaonic atoms X-ray spectroscopy.\\
Two types of background are considered, electromagnetic and hadronic. The electromagnetic one, asynchronous with the kaons' production, is generated by particles lost by DA$\Phi$NE's circulating beams due to the beam-gas interaction and the Touschek effect.\\
The kaon trigger (KT), consisting of two plastic scintillators placed above and below the interaction region, is used to detect the back-to-back emitted K$^+$K$^-$ pairs. The coincidence between the two scintillators provides the trigger signal that allows to reject the hits on the SDDs not synchronous with kaons.\\
The hadronic background is related to the K$^+$ decay and the K$^-$ nuclear absorption resulting in the emission of particles (MIPs), mostly pions and muons, releasing a signal in the SDDs synchronous with the KT signal. To overcome this drawback, three different veto systems \cite{Bazzi:2013kwa}, placed behind the SDDs and around the vacuum chamber, are employed to detect and reject the MIP-induced signals. The setup is also equipped with a luminosity monitor \cite{Skurzok:2020phi}, placed on the longitudinal plane in front of the IR, to monitor the background and measure the collider luminosity in real-time.\\
In 2022 the SIDDHARTA-2 setup was installed on DA$\Phi$NE and then optimized by performing kaonic helium transitions measurements to the 2p level, which have a much higher yields than the transitions on the 1s level in kaonic deuterium. The target cell was filled with helium-4 at the density of 1.37 $\pm$ 0.07 g/l (1.1\% liquid helium density). No veto systems were installed at that time; for this reason only the electromagnetic background reduction procedure is applied in this work.

\section{Data selection}
\begin{figure}[htbp]
\centering
\mbox{\includegraphics[width=15 cm]{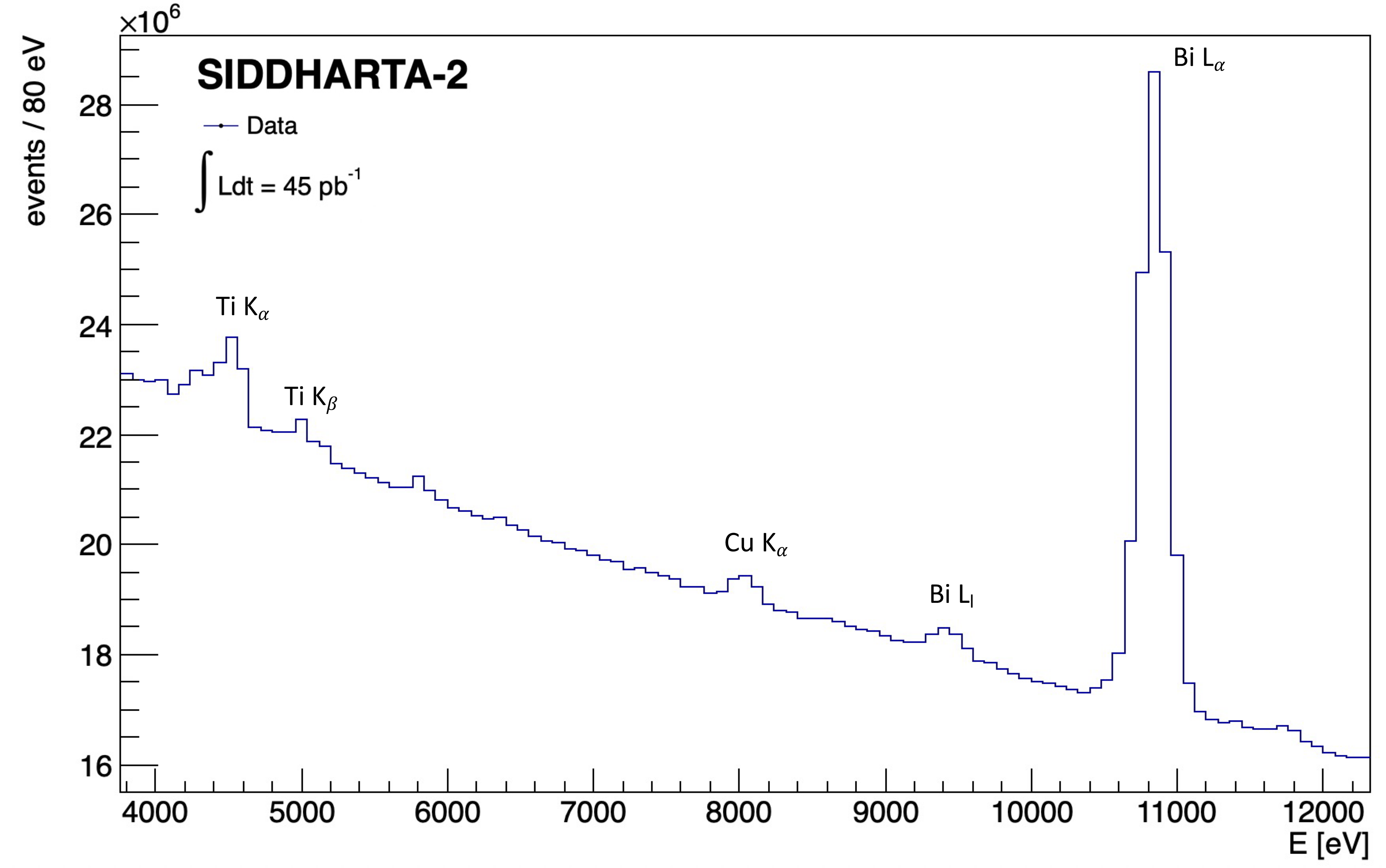}}
\caption{Inclusive kaonic helium-4 energy spectrum.}
\label{fig_datanotrig}
\end{figure}

\begin{figure}[htbp]
\centering
\mbox{
\includegraphics[width=8.5 cm]{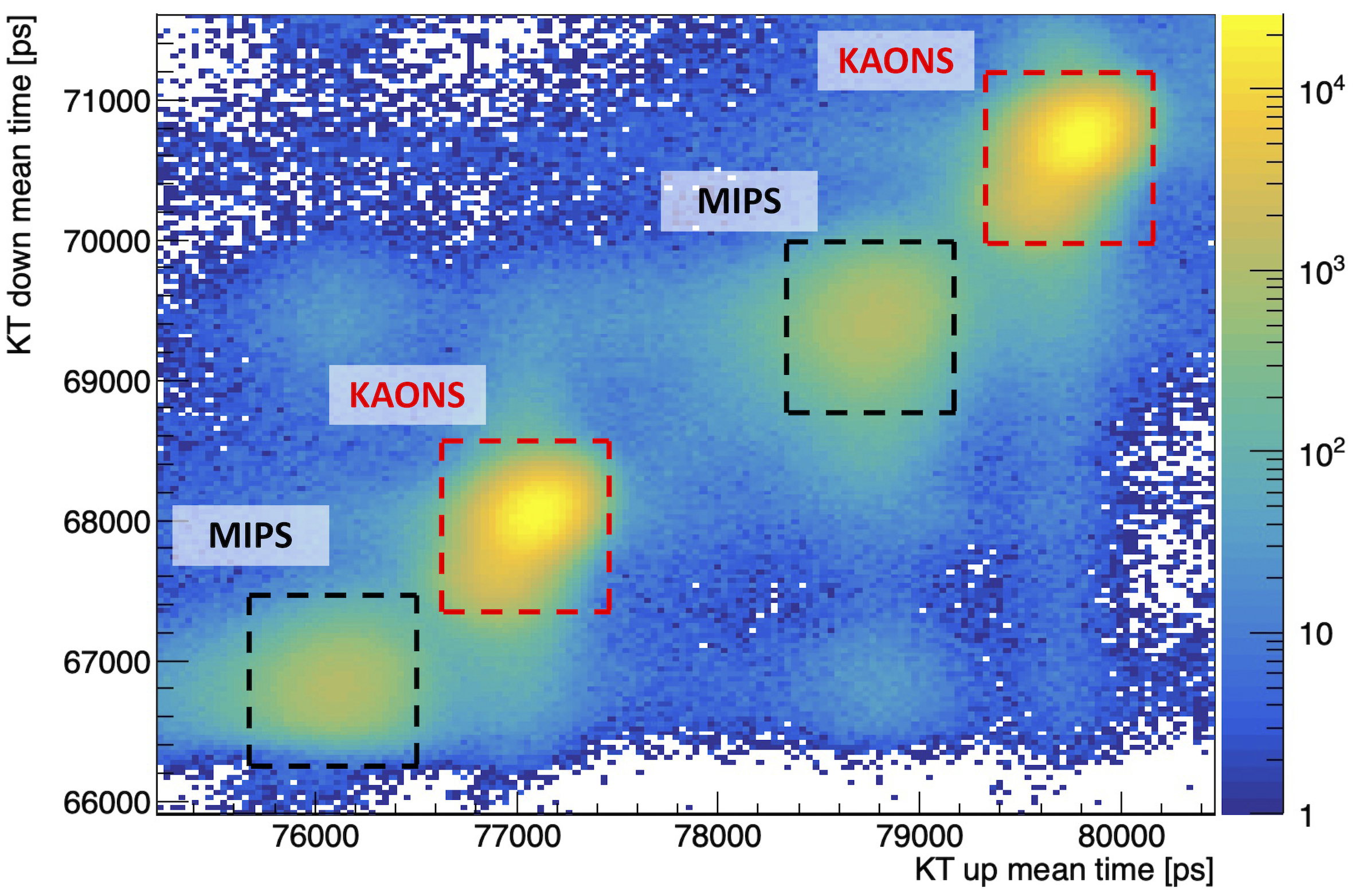}
\includegraphics[width=7.5 cm]{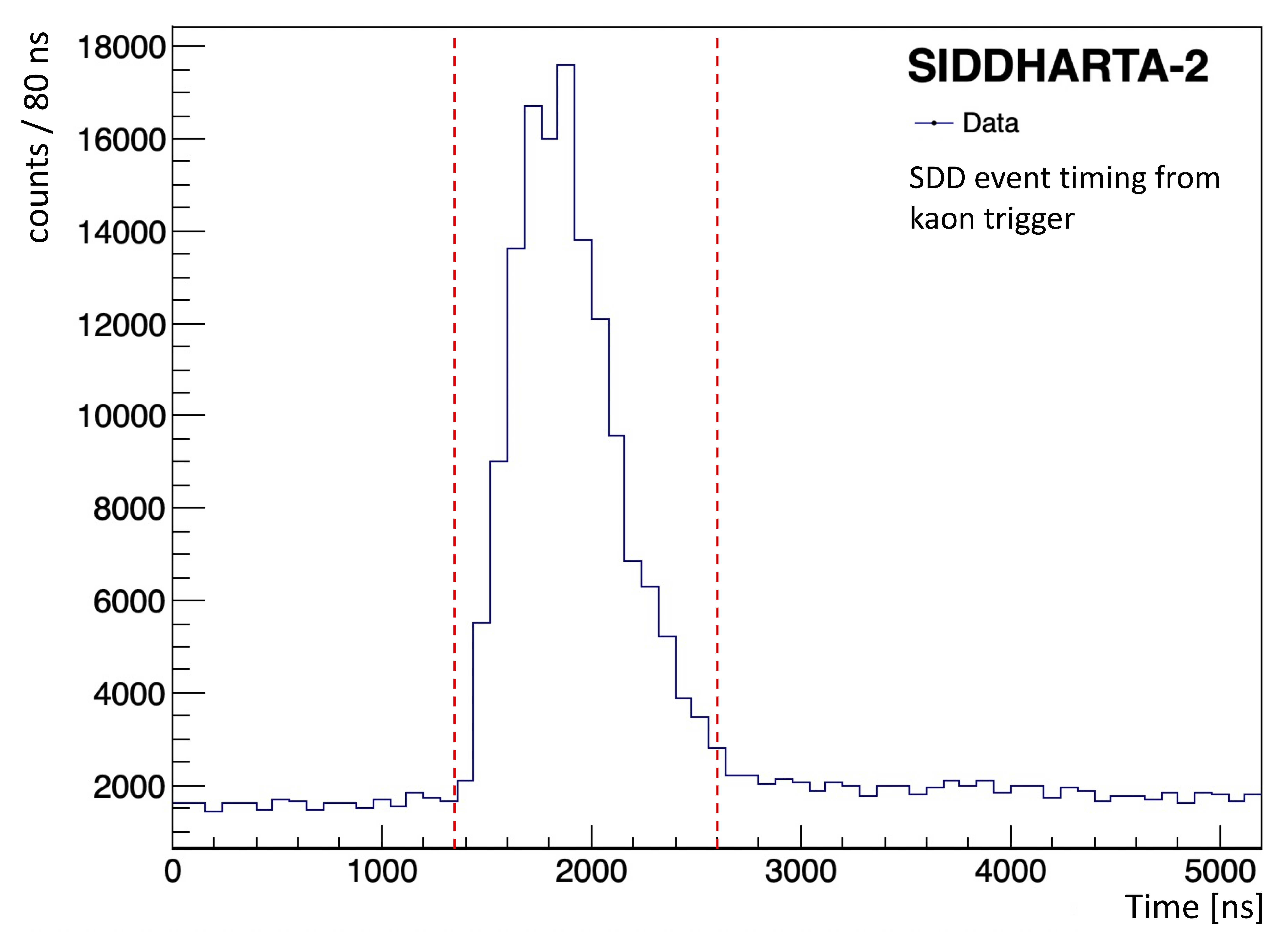}
}
\caption{Left: Two-dimensional scatter plot of KT time distributions. The coincidence events related to kaons (high intensity) are clearly distinguishable from MIPs (low intensity). Right: Time difference between the KT signals and X-ray hits on the SDDs. The dashed lines represent the acceptance window.}
\label{fig_kaon_drift}
\end{figure}

\begin{figure}[htbp]
\centering
\mbox{\includegraphics[width=15 cm]{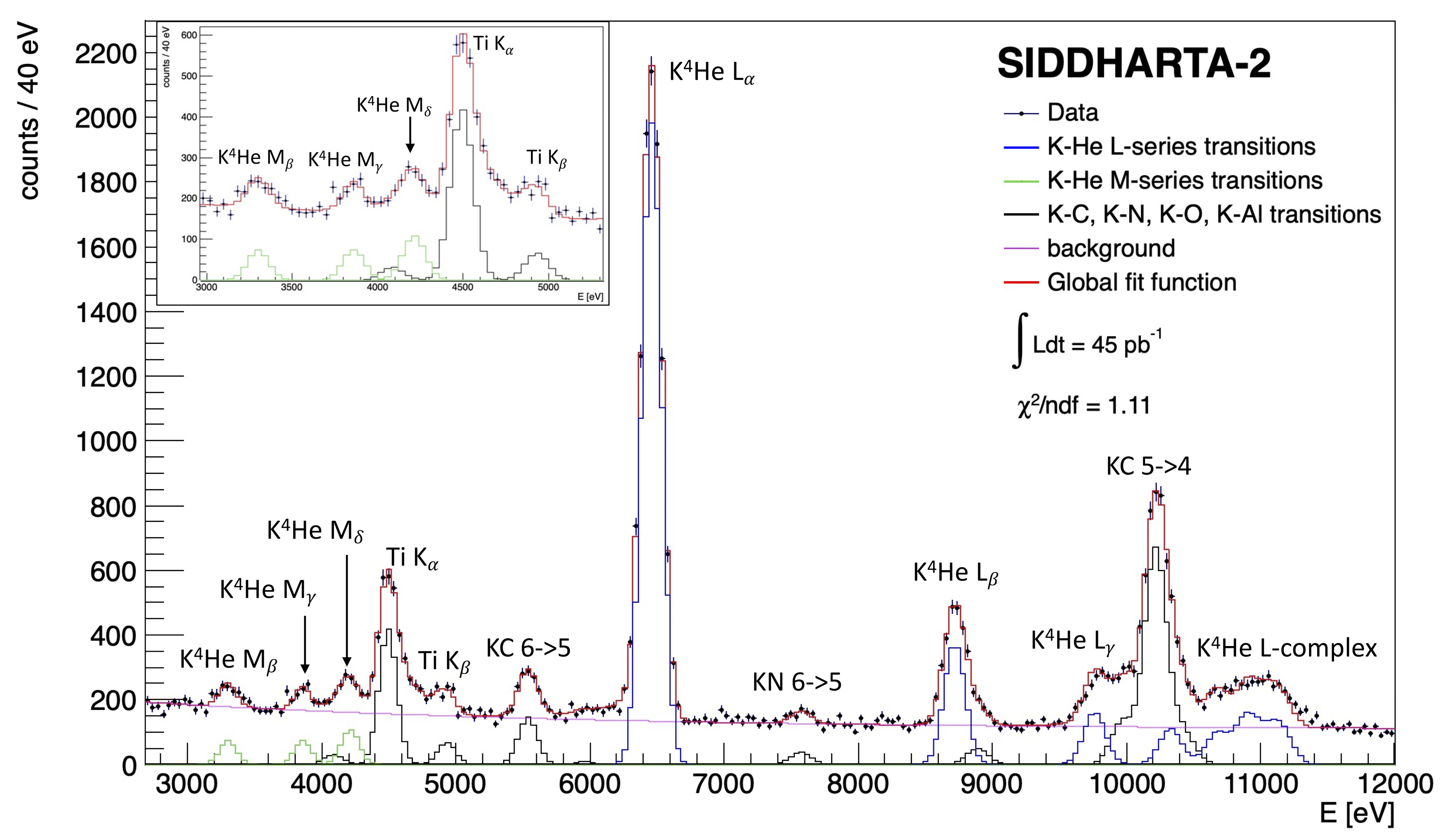}}
\caption{X-ray energy spectrum and fit of the data after the background suppression procedure (see text). The kaonic helium-4 L-series and M-series transitions are indicated. }
\label{fig_fit}
\end{figure}

Figure \ref{fig_datanotrig} shows the inclusive energy spectrum acquired by the SDDs during the helium-4 run for a total integrated luminosity of 45 pb$^{-1}$. The spectrum displays fluorescence peaks which correspond to the X-ray emission of materials placed around the SDDs. Titanium and copper lines are from setup components inside the vacuum chamber, while the bismuth comes from the alumina ceramic boards behind the SDDs.\\
The high continuous background contribution prevents to directly observe the kaonic helium signal. In this context, the KT plays a crucial role. Only the events falling in a 5 $\mu$s time window in coincidence with a trigger signal are selected, rejecting a substantial fraction of the background. The time window width was tuned to enable the front-end electronics to process and acquire the signals. However, there are cases where MIPs, generated by beam-beam and beam-gas interactions, can produce a trigger signal when they simultaneously pass through the KT scintillators. To distinguish between these MIP-induced triggers and those originating from K$^+$K$^-$ pairs, a Time of Flight (TOF) analysis is employed. This technique relies on measuring the temporal difference between the trigger signal and the DA$\Phi$NE radio-frequency (RF), which serves as a collision reference. Figure \ref{fig_kaon_drift} shows the mean time distribution measured by the two KT scintillators and the TOF cut used to reject the MIP-induced triggers.\\
In order to enhance the background rejection, the time difference between the KT signal and the time of X-ray detection was evaluated. This time distribution is shown in Figure \ref{fig_kaon_drift}; the main peak within the red dashed lines corresponds to hits on the SDDs in coincidence with the trigger, while the flat distribution is given by uncorrelated events. The combined use of KT and SDDs time information allowed to reduce the background by a factor $\sim$10$^5$, resulting in the final energy spectrum shown in Figure \ref{fig_fit}.

\section{Data analysis, results and discussion}
After the selection procedure, the kaonic helium-4 M-series and L-series lines are clearly visible (Figure \ref{fig_fit}) in the energy regions 3.0 keV - 4.5 keV and 6.4 keV - 12 keV, respectively. Additionally, the X-rays lines corresponding to kaonic carbon, oxygen, nitrogen and aluminium, generated by kaons stopped in the apparatus support frame and in the kapton window of the target cell, are also detected. The measurements and a detailed investigation of these lines are reported in \cite{Sgaramella:2023orc}.\\
The kaonic helium-4 peaks were fitted to extract their energies and the number of events associated to each line. The detector energy response function is given by the convolution of a Gaussian (Eq. \ref{gauss_func}) with a tail-function (Eq. \ref{tail_func}) \cite{Gysel:2003}, used to account for the electron-hole recombination and the incomplete charge collection effect. The width ($\sigma$) of the Gaussian represents the energy resolution of the SDDs, and is described as a function of three parameters: the Fano Factor (FF) \cite{Fano:1947zz}, the electron-hole pair energy creation ($\epsilon$), the electronic and thermal noises (noise).\\
\begin{equation}
     G(x)=\frac{A_G}{\sqrt{2\pi}\sigma} \cdot e^{\frac{-(x-x_0)^2}{2\sigma^2}} \\
     \sigma=\sqrt{FF\cdot\epsilon\cdot E+\frac{noise^2}{2.35^2}} 
     \label{gauss_func}
\end{equation}
\begin{equation}
    T(x)=\frac{A_T}{2\beta\sigma} \cdot e^{\frac{x-x_0}{\beta\sigma}+\frac{1}{2\beta^2}} \cdot\emph{erfc}\left(\frac{x-x_0}{\sqrt{2}\sigma}+\frac{1}{\sqrt{2}\beta}\right)
    \label{tail_func}
\end{equation}
The $A_G$ and $A_T$ are the amplitudes of the Gauss and Tail functions, respectively, while the $\beta$ parameter is the slope of the tail; the \emph{erfc} term stands for the complementary error function.\\
Since the 2p level could be affect by the strong interaction between kaon and nucleus, a Lorentzian function (Eq. \ref{lorentz_func}), accounting for the intrinsic line width ($\Gamma$) induced by the strong interaction, convoluted with the Gaussian and the tail-function, was used to fit the L-series peaks.\\
\begin{equation}
    L(x)=\frac{1}{\pi} \frac{\frac{1}{2}\Gamma}{\left( x-x_0 \right)^2 + \left( \frac{1}{2}\Gamma \right)^2 }
    \label{lorentz_func}
\end{equation}
For a full description of the spectrum shape, an exponential function plus a first degree polynomial were employed to reproduce the continuous background below the peaks. The global fit function, shown in Figure \ref{fig_fit}, properly reproduces the data distribution in the energy range from 3 to 12 keV, with a $\chi^2 / ndf$ = 1.11.

\subsection{Kaonic helium-4 M-series transitions}
Using the gaseous target allowed to observe and measure, for the first time, M-lines in kaonic helium, in particular the M$_\beta$, M$_\gamma$, and M$_\delta$ transitions. Their energies are reported in Table \ref{tab_energy}. The associated systematic errors were calculated taking into account the linearity and stability of the SDDs, as well as the calibration accuracy \cite{Sgaramella:2022}. The M$_\alpha$ transition was not observed because, having an energy lower than 3 keV, it is absorbed by the target cell kapton walls.

\begin{table}[ht]
\caption{The measured energies of the kaonic helium-4 L$_\alpha$, M$_\beta$, M$_\gamma$, and M$_\delta$ transitions.}
\label{tab_energy}
\centering
\begin{tabular}{@{}llc}
\br
Transition & X-ray name & Energy \\
\mr
3d$\rightarrow$2p & \textit{L$_\alpha$} & $\mathrm{6461.4\pm 0.8 \,(stat) \pm 2.0 \,(sys)\ eV}$ \\
5f$\rightarrow$3d & \textit{M$_\beta$} & $\mathrm{3300.8\pm 13.2 \,(stat) \pm 2.0 \,(sys)\ eV}$ \\
6f$\rightarrow$3d & \textit{M$_\gamma$} & $\mathrm{3860.4\pm 13.6 \,(stat) \pm 2.2 \,(sys)\ eV}$ \\
7f$\rightarrow$3d & \textit{M$_\delta$} & $\mathrm{4214.1\pm 19.6 \,(stat) \pm 2.2 \,(sys)\ eV}$ \\
\br
\end{tabular}
\end{table}

\subsection{A new kaonic helium-4 L$_\alpha$ transition measurement}
Assuming the effect of the strong interaction on the 3d level to be negligible, the energy shift ($\epsilon_{2p}$) has been extracted for the 2p level from the difference between the measured L$_\alpha$ transition energy ($\mathrm{E^{exp}_{3d\rightarrow2p}}$), reported in Table \ref{tab_energy}, and the electromagnetic value ($\mathrm{E^{e.m}_{3d\rightarrow2p}}$) calculated by considering vacuum polarization and the recoil effect \cite{Santos:2004bw}. The width ($\Gamma_{2p}$) is directly derived from the $\Gamma$ parameter of the Lorentzian function used to fit the L$_\alpha$ peak. Therefore, the measured strong interaction induced shift and width of the 2p level in kaonic helium-4 are:
\begin{eqnarray}
&\epsilon_{2p}= \mathrm{E^{exp}_{3d\rightarrow2p} - E^{e.m}_{3d\rightarrow2p}} = \mathrm{-1.9\pm 0.8 \,(stat) \pm 2.0 \,(sys)\ eV}& \\
&\Gamma_{2p} = \mathrm{0.01\pm 1.60 \,(stat) \pm 0.36 \,(sys)\ eV}&
\end{eqnarray}
The systematic uncertainty on the shift is related to the accuracy of the SDDs calibration, whereas the one on the width is given by the inaccuracy on the SDDs energy resolution.\\
The reported results show that there is no sharp effect of the strong interaction on the 2p level, confirming the past measurements on gaseous kaonic helium-4, and improving by a factor of three the statistical precision on the 2p level shift and width, making it the most precise measurement in a gas target. 

\subsection{The L and M-series transitions X-ray yields}
\begin{figure}[htbp]
\centering
\mbox{\includegraphics[width=10 cm]{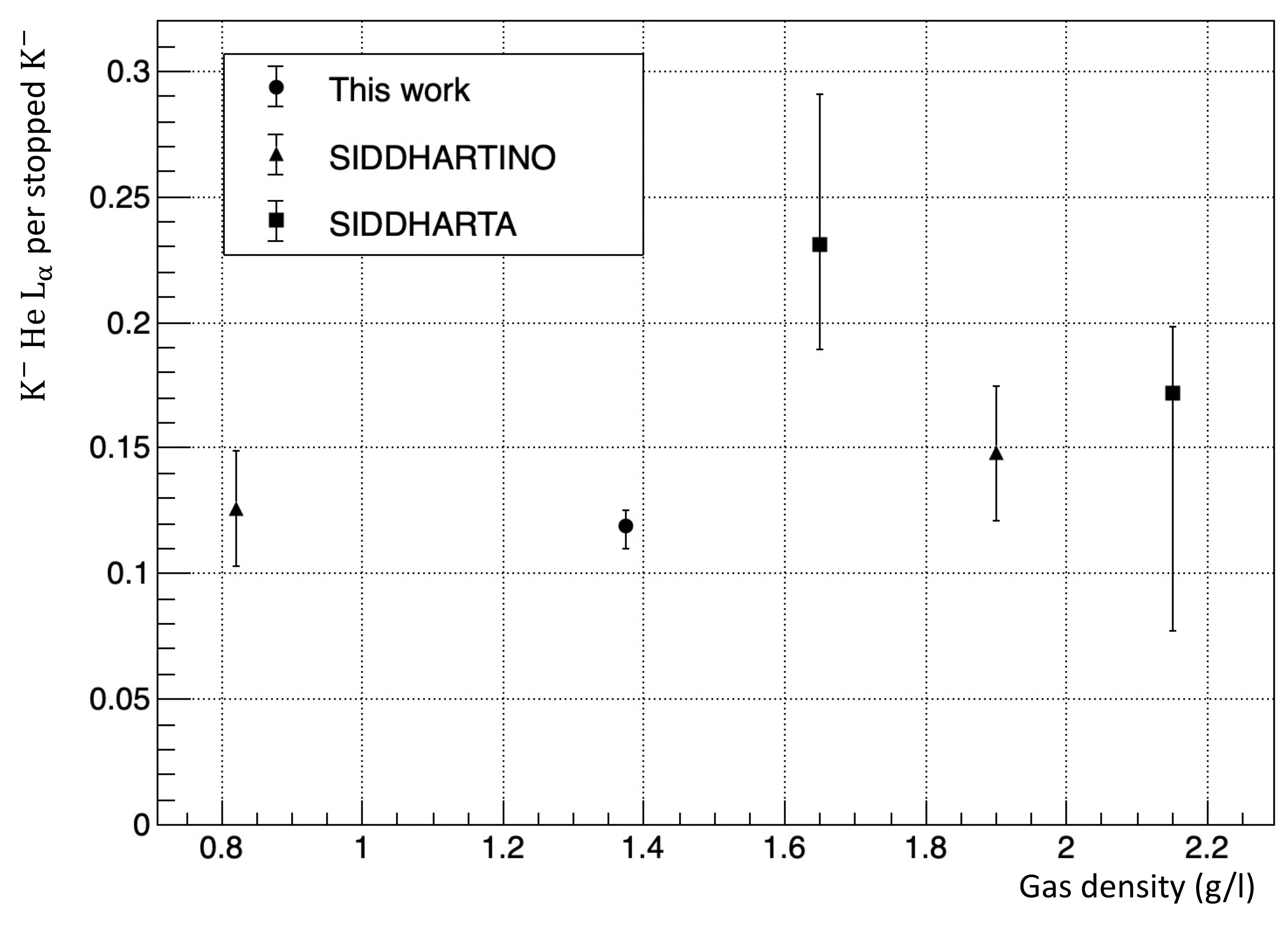}}
\caption{K$^-$ $^4$He L$_{\alpha}$ X-ray yield as function of the target density from this work, SIDDHARTINO \cite{Sirghi:2023scw}, and SIDDHARTA \cite{Bazzi:2014}.}
\label{fig_Layield}
\end{figure}

Monte Carlo simulations are used to evaluate the fraction of kaons stopping in the gas targets, which is necessary to extract the absolute X-ray yields. The absolute yield (Y) for an X-ray transition per stopped kaon is given by the ratio between the experimental detection efficiency ($\epsilon^{EXP}$) and the Monte Carlo efficiency ($\epsilon^{MC}$). The Monte Carlo simulation code is based on the GEANT4 toolkit, where all the materials and geometries used in the experiment were included.
The $\epsilon^{EXP}$ is obtained by normalizing the number of measured X-rays ($N_{X-ray}^{exp}$) to the number of kaon triggers ($N_{KT}^{exp}$) and the active area of the detectors. Similarly, the $\epsilon^{MC}$ is given by the number of simulated X-rays ($N_{X-ray}^{MC}$), normalized to the number of simulated kaon triggers ($N_{KT}^{MC}$). The kaonic atom X-rays were generated at the position where the $K^{-}$ stopped, and were isotropically emitted with a 100$\%$ yield for each transition and each kaonic atom. Hence, the absolute X-ray yield is given by:
\begin{equation}
Y = \frac{\epsilon^{EXP}}{\epsilon^{MC}} = \frac{N_{X-ray}^{exp}/N_{KT}^{exp}}{N_{X-ray}^{MC}/N_{KT}^{MC}}
\label{eq_yield}
\end{equation}
\noindent
We extracted the number of events for each kaonic helium-4 transition from the fit of the energy spectrum (Figure \ref{fig_fit}), to evaluate the X-ray yields of the L and M-series transitions. The number of events for each transition is listed in Table \ref{tab_events} with the statistical uncertainty given by the fit. The absolute yields for the kaonic helium-4 L$_\alpha$ and M$_\beta$ transitions were obtained by applying Eq. (\ref{eq_yield}) and are reported in Table \ref{tab_yield} with their statistical and systematic uncertainties. The relative yields of the L$_\beta$, L$_\gamma$, M$_\gamma$, and M$_\delta$ transitions, taking into account the energy-dependent detection efficiency, were also evaluated and are reported in Table \ref{tab_yield}.\\
\begin{table}[htbp]
\caption{Number of events for the kaonic helium-4 L-series and M-series transitions, obtained by the fit of the spectrum in Figure \ref{fig_fit}.}
\label{tab_events}
\centering
\begin{tabular}{@{}lll}
\br
Transition & X-ray name &number of events\\
\mr
3d$\rightarrow$2p & \textit{L$_\alpha$} & $\mathrm{9158\pm 133 }$ \\
4d$\rightarrow$2p & \textit{L$_\beta$} & $\mathrm{1852\pm 62 }$ \\
5d$\rightarrow$2p & \textit{L$_\gamma$} & $\mathrm{139\pm 9 }$ \\
5f$\rightarrow$3d & \textit{M$_\beta$} & $\mathrm{289\pm 36 }$ \\
6f$\rightarrow$3d & \textit{M$_\gamma$} & $\mathrm{306\pm 33 }$ \\
7f$\rightarrow$3d & \textit{M$_\delta$} & $\mathrm{365\pm 55 }$ \\
\br
\end{tabular}
\end{table}
\noindent
The main source of systematic uncertainty for the yields measurements is related to the accuracy with which the gas density is known. The gas density is a key input for the Monte Carlo simulation, since it affects the number of kaons stopped in the gas target, and consequently the number of X-ray events. The helium gas density was determined by measuring the gas pressure and temperature. The uncertainties induced by the temperature and pressure sensors are $\pm$2\% and $\pm$3.5\%, respectively, leading to a density error of $\pm$5\%. Monte Carlo simulations were used to estimate the systematic error on the absolute yields due to uncertainty on the gas density. Instead, for the relative yields the systematic error is negligible with respect to the statistical one. \\
It is worth to underline that the results shown in Table \ref{tab_yield} represent the first experimental measurement of the M-series transitions yields, providing new experimental data to optimize the cascade models for kaonic helium and, more generally, kaonic atoms. Furthermore, the measurement of the L$_\alpha$ X-ray yield at the density of 1.37 $\pm$ 0.07 g/l establishes a new experimental record and data point that, combined with the measurements performed by SIDDHARTA \cite{Bazzi:2014} and SIDDHARTINO \cite{Sirghi:2023scw}, will allow to check and improve kaonic atoms cascade models across the density scale (see Figure \ref{fig_Layield}).

\begin{table}[htbp]
\caption{The absolute yields of the kaonic helium-4 L$_\alpha$ and M$_\beta$ transitions and the relative yields of L$_\beta$, L$_\gamma$, M$_\gamma$, and M$_\delta$ transitions}
\label{tab_yield}
\centering
\begin{tabular}{@{}ll}
\br
Density& 1.37 $\pm$ 0.07 g/l\\
\mr
\textit{L$_\alpha$ yield} & $ \mathrm{ 0.119 \pm 0.002 \,(stat) ^{+0.006\,(sys)}_{-0.009\,(sys)} } $ \\
\textit{M$_\beta$ yield} & $ \mathrm{ 0.026 \pm 0.003 \,(stat) ^{+0.010\,(sys)}_{-0.001\,(sys)} } $ \\
\mr
\textit{L$_\beta$ / L$_\alpha$} & $ \mathrm{ 0.172 \pm 0.008 \,(stat) } $ \\
\textit{L$_\gamma$ / L$_\alpha$} & $ \mathrm{ 0.012 \pm 0.001 \,(stat) } $ \\
\textit{M$_\beta$ / L$_\alpha$} & $ \mathrm{ 0.218 \pm 0.029 \,(stat) } $ \\
\textit{M$_\gamma$ / M$_\beta$} & $ \mathrm{ 0.48 \pm 0.11 \,(stat) } $ \\
\textit{M$_\delta$ / M$_\beta$} & $ \mathrm{ 0.43 \pm 0.12 \,(stat) } $ \\
\br
\end{tabular}
\end{table}

\newpage
\section{Conclusions}
In this work, we presented a comprehensive investigation of kaonic helium-4 through X-ray spectroscopy. The excellent SDD energy response and remarkable SIDDHARTA-2's background suppression, allowed to detect and measure, for the first time, the energies of three M-series transitions. Our measurement of the X-ray yields for several M-series and L-series transitions in kaonic helium-4, provide new fundamental input to cascade models, potentially contributing to a deeper understanding of the de-excitation mechanisms within kaonic atoms. The measurement of the L$_\alpha$ yield at the density of 1.37 $\pm$ 0.07 g/l, combined with the previous results obtained by SIDDHARTA \cite{Bazzi:2014} and SIDDHARTINO \cite{Sirghi:2023scw}, offers a new opportunity to understand the density dependence of the kaonic helium yield. Moreover, the measurement of the 2p level energy shift and width improves the statistical accuracy by a factor three, compared to the previous results with gaseous helium-4, definitely rejecting the hypothesis of a large energy shift and width. \\
The SIDDHARTA-2 outcome refines our understanding of the kaonic helium-4 system, and contribute to the ongoing efforts to comprehend the strong interaction in the non-perturbative regime in systems with strangeness.

\section*{Acknowledgments}
We thank C. Capoccia from LNF-INFN and H. Schneider, L. Stohwasser, and D. Pristauz-Telsnigg from Stefan Meyer-Institut for their fundamental contribution in designing and building the SIDDHARTA-2 setup. We thank as well the DA$\Phi$NE staff for the excellent working conditions and permanent support. Part of this work was supported by the Austrian Science Fund (FWF): [P24756-N20 and P33037-N] and FWF Doctoral program No. W1252-N27; the EXOTICA project of the Minstero degli Affari Esteri e della Cooperazione Internazionale, PO22MO03; the Croatian Science Foundation under the project IP-2018-01-8570; the EU STRONG-2020 project (Grant Agreement No. 824093); the EU Horizon 2020 project under the MSCA (Grant Agreement 754496); the Japan Society for the Promotion of Science JSPS KAKENHI Grant No. JP18H05402; the Polish Ministry of Science and Higher Education grant No. 7150/E-338/M/2018 and the Polish National Agency for Academic Exchange( grant no PPN/BIT/2021/1/00037); the EU Horizon 2020 research and innovation programme under project OPSVIO (Grant Agreement No. 101038099). The authors acknowledge support from the SciMat and qLife Priority Research Areas budget under the program Excellence Initiative—Research University at the Jagiellonian University. Catalina Curceanu acknowledge University of Adelaide, where part of this work was done (under the George Southgate fellowship, 2023).

\section*{References}
\bibliography{iopart-num}

\end{document}